\documentclass{amsart}

\usepackage{graphicx}
\usepackage{color}
\newtheorem{theorem}{Theorem}%[section]
 \newtheorem{lem}{Lemma}%[section]

  \newtheorem{remark}{Remark}%[section]
\newtheorem{corollary}{Corollary}
\UseRawInputEncoding
\usepackage[russian, english]{babel}

\def\dfrac#1#2{\displaystyle{#1\over #2}}

\def\zo{\omega}

\def\bv{{\bf v}}
\def\bV{{\bf V}}

\def\Div{\mbox{div}\,}

\def\bB{{\bf B}}

\def\bE{{\bf E}}
\begin{document}

\markboth{Delova, Rozanova}{The Interplay of Regularizing Factors}
% in the Model of Upper Hybrid Oscillations}

%%%%%%%%%%%%%%%%%%% Publisher's Area please ignore %%%%%%%%%%%%%%%%%%%%%%%
%
%\catchline{}{}{}{}{}
%
%%%%%%%%%%%%%%%%%%%%%%%%%%%%%%%%%%%%%%%%%%%%%%%%%%%%%%%%%%%%%%%%%%%%%%%%%%

\title[Regularizing Factors in the Model of Upper Hybrid Oscillations]{
%ON ELECTROSTATIC OSCILLATIONS IN COLD MAGNETOACTIVE PLASMA\\
The Interplay of Regularizing Factors in the Model of Upper Hybrid Oscillations of Cold Plasma
%Interaction of Regularizing Factors in the Model of Upper Hybrid Oscillations of Cold Plasma
}

\author{Maria I. Delova}

\address{ Mathematics and Mechanics Department, Lomonosov Moscow State University, Leninskie Gory,
Moscow, 119991,
Russian Federation,
mashadelova@yandex.ru}

\author{Olga S. Rozanova*}

\address{ Mathematics and Mechanics Department, Lomonosov Moscow State University, Leninskie Gory,
Moscow, 119991,
Russian Federation,
rozanova@mech.math.msu.su}

\subjclass{Primary 35Q60; Secondary 35L60, 35L67, 34M10}

\keywords{Quasilinear hyperbolic system,
plasma oscillations, electron-ion collisions, magnetic effect, blow up}

\maketitle

%\begin{history}
%\received{(Day Month Year)}
%\revised{(Day Month Year)}
%\accepted{(Day Month Year)}
%\comby{(xxxxxxxxxx)}
%\end{history}

\begin{abstract}
 A one-dimensional nonlinear model of the so-called upper hybrid oscillations in a magnetoactive plasma is investigated taking into account electron-ion collisions. It is known that both the presence of an external magnetic field of strength $ B_0 $ and a sufficiently large collisional factor $ \nu $ help suppress the formation of a finite-dimensional singularity in a solution (breaking of oscillations). Nevertheless, the suppression mechanism is different: an external magnetic field increases the oscillation frequency, and collisions tend to stabilize the medium and suppress oscillations.
 In terms of the initial data and the coefficients $ B_0 $ and $ \nu $, we establish a criterion for maintaining the global smoothness of the solution.
Namely, for fixed $ B_0 $ and $ \nu \ge 0 $
 one can precisely divide the initial data into two classes: one leads to stabilization to the equilibrium, and the other leads to the destruction of the solution in a finite time.

 Next, we examine the nature of the stabilization. We show that for small $ |B_0| $ an increase in the intensity factor first leads to a change in the oscillatory behavior of the solution to monotonic damping, which is then again replaced by oscillatory damping. At large values of $ |B_0| $, the solution is characterized by oscillatory damping regardless of the value of the intensity factor $ \nu $.
 \end{abstract}

%\keywords{quasilinear hyperbolic system;
%plasma oscillations; breaking effect; loss of smoothness.}

%\ams{AMS Subject Classification: 35Q60, 35L60, 35L67, 34M10}
\section*{Introduction}

\bigskip

The equations of hydrodynamics of "cold" plasma  in the non-relativistic approximation take the form
\begin{equation}
\
\label{base1}
\begin{array}{l}

\dfrac{\partial n }{\partial t} + \Div(n \bV)=0\,,\quad
\dfrac{\partial \bV }{\partial t} + \left( \bV \cdot \nabla \right) \bV
=\frac {e}{m} \, \left( \bE + \dfrac{1}{c} \left[\bV \times  \bB\right]\right) - \nu_e \bV,\vspace{0.5em}\\
\dfrac1{c} \frac{\partial \bE }{\partial t} = - \dfrac{4 \pi}{c} e n \bV
 + {\rm rot}\, \bB\,,\quad
\dfrac1{c} \frac{\partial \bB }{\partial t}  =
 - {\rm rot}\, \bE\,, \quad \Div \bB=0\,,
\end{array}
\end{equation}
where
$ e, m $ --- charge and mass of the electron (here the electron charge has a negative sign: $ e <0 $),
$ c $ --- speed of light;
$ n,  \bV $ --- density and velocity of
electrons;
$ \bE, \bB $ --- vectors of electric and magnetic fields,
the term $- \nu_e \bV $ describes electron-ion collisions, which
can be interpreted as a friction between particles,  see, for example,~\cite{ABR78}. System \eqref{base1} is described in sufficient detail in textbooks and monographs (see, for example,   \cite {GR75}).
Currently, the great attention is paid to a study of the cold plasma due to a possibility of acceleration of electrons in the wake wave of a powerful laser pulse  \cite {esarey09}, nevertheless theoretical results in this field are very scarce.

It is commonly known that the plasma oscillations described by \eqref{base1}, tend to break. Mathematically,
 the breaking process means a blow up  of solution, and the appearance of a delta-shape singularity of the electron density  \cite {Dav72}. Among main interests  is a study of possibility of existence of a smooth solution as long as possible.

 To obtain  exact mathematical results we use the model of {\it upper hybrid} oscillations of cold magnetoactive plasma (see, for example,  \cite {Dav72}),  a simplification that the
 physicists traditionally use to study the effect of an external magnetic field. The model  was appeared, apparently, for the first time, in \cite{Dav68}, further it was used in various contexts, e.g.  \cite{Maity12}, \cite{Maity13_PRL}, \cite{Kar16}. To derive the model they expand the solution to \eqref{base1} in a series with respect a small parameter and retain physically significant terms.
In the Cartesian coordinates $x, y, z$ the solution has the following structure:
 $$
 \bV=(V_x(x,t), V_y(x,t), 0),\,\, \bE(x,t) = (E_x(x,t),0,0), \,\, \bB(x,t) = (0,0,b_0), \, b_0 \equiv {\rm const}.
$$
In  the dimensionless quantities
$$
\begin{array}{c}
\rho = k_p x, \quad \theta = \omega_p t, \quad
{V_1} = \dfrac{V_x}{c}, \quad
{V_2} = \dfrac{V_y}{c}, \vspace{1 ex}\\
{E} = -\,\dfrac{e\,E_x}{m\,c\,\omega_p}, \quad
{ N} = \dfrac{n}{n_0}, \quad
{ B}_0 = -\,\dfrac{e\,b_0}{m\,c\,\omega_p},\quad \nu = \dfrac{\nu_{e}}{\omega_p},
\end{array}
$$
where $ \omega_p = \left (4 \pi e^ 2 n_0 / m \right)^{1/2} $ is the plasma frequency, $ n_0 $ is the value of the unperturbed electron
density, $ k_p = \omega_p / c $, the result takes the form
\begin{equation}
\label{3gl5non}
\begin{array}{c}
\dfrac{\partial V_1 }{\partial \theta}+
V_1 \dfrac{\partial V_1}{\partial \rho}  = -  E- B_0V_2-\nu V_1, \quad \\
\dfrac{\partial V_2 }{\partial \theta}+
V_1 \dfrac{\partial V_2}{\partial \rho}  =  B_0V_1-\nu V_2, \quad \\
\dfrac{\partial  E }{\partial \theta}+V_1\dfrac{\partial  E}{\partial \rho} = V_1.
\end{array}
\end{equation}
The solutions of this system are called in \cite{Dav68}
nonlinear zero-temperature Bernstein modes.

We consider (\ref {3gl5non}) with the initial conditions
\begin{equation}\label{cd2}
     V_1(\rho,0) = V_1^0(\rho), \quad V_2(\rho,0) = V_2^0(\rho), \quad
     E(\rho,0) = E^0(\rho), \quad
 \rho \in {\mathbb R}.
\end{equation}

 Due to the Gauss theorem the electron density can be found as
\begin{equation}
 {N}(\rho,\theta) = 1 -
\dfrac{\partial  {E}(\rho,\theta) }{\partial \rho}.
\label{3gl4den}
\end{equation}

System  (\ref {3gl5non}) is of hyperbolic type. It is well known that for such systems, there exists, locally in time, a unique solution to the Cauchy problem of the same class as the initial data. % in our case it is $ C^1 $.
The blow up of a solution is associated with unbounded derivatives \cite {Daf16}.

The Cauchy problem \eqref {3gl5non}, \eqref{cd2} was studied in the following cases:
\begin{itemize}
\item $\nu=B_0=0$ in \cite{RChZAMP21};
\item $B_0=0$, $\nu>0$ in \cite{RChD20}, see also \cite{RChZAMP21_2} for the relativistic case;
\item $\nu=0$, $B_0\ne 0$ in \cite{bib:03}.
\end{itemize}
In all these case we obtained the criterion of the singularity formation for the solution in terms of the initial data and parameters $\nu$, $B_0$. In other words, for any specific initial data we can say in advance whether the solution keeps the initial smoothness
 for all $\theta>0$ or blows up in a finite time $\theta_*>0$. Moreover, for the case $\nu>0$ the global smoothness of solution means a stabilization to the trivial equilibrium state \cite{RChZAMP21_2}.

 Both parameters $\nu$ and $B_0$ act as  regularizers in the following sense: if we fix an arbitrary initial data  \eqref{cd2}, we can obtain the global smoothness of solution by increasing $\nu$ or $|B_0|$. For example, the criterion of the singularity formation for the case $\nu=0$ is the following.
 \begin{theorem} \label{T1} \cite{bib:03}. For the existence of a $ C^1 $ -- smooth $ \frac {2 \pi} {\sqrt {1 + B_0^2}} $ - periodic solution \\ $ (V_1 (\theta, \rho), \, V_2 (\theta, \rho), \, E (\theta, \rho)) $ of the problem  (\ref {3gl5non}) with $\nu=0$, (\ref {cd2}), with  $(V_1^0, V_2^0, E^0)\in C^2({\mathbb R}) $, it
is necessary and sufficient that at any point $ \rho \in \mathbb R $
 \begin{equation}\label {crit2}
\left( (V_1^0)' \right) ^ 2 + 2 \, (E^0)' +2 B_0\, (V_2^0)' - B_0^2  -1 <0.
\end {equation}
If the opposite inequality (\ref{crit2}) holds at least at one point $ \rho_0 $, then the derivatives of the solution turn to infinity in a finite time.
\end{theorem}
We can see that the frequency of oscillations increases with $B_0$ and the smooth solution never stabilazes to a constant state.

For $ B_0 = 0 $ we can draw on the plane $((E^0)'( \rho_0), (V_1^0)'(\rho_0))$ a smooth curve dividing the plane into two domains: one of them corresponds to a globally smooth solution that stabilizes to the zero equilibrium if the point $((E^0)'( \rho_0), (V_1^0)'(\rho_0))$ gets there for all $\rho_0\in \mathbb R$. If, for certain initial data, the point $((E^0)'( \rho_0), (V_1^0)'(\rho_0))$ does not belong to this domain for at least one $ \rho_0 $, the corresponding solution  blows up. The domain of smoothness expands with increasing $ \nu $, and the character of stabilization changes with  $ \nu $: for $ \nu \in (0,2) $ the damping is oscillatory, while for $ \nu \ge 2 $ damping is monotonic starting from a certain moment of time.

The purpose of this article is to study the relationship between the regularizing factors $ \nu $ and $ B_0 $. It is natural to expect that a smooth solution is stabilized in the presence of both factors, and the high intensity of electron-ion collisions suppresses oscillations. However, as we are going to show, the nature of the stabilization is drastically different. Namely, for any fixed $ B_0 $ there is a limit of the oscillation frequency $ \Omega (B_0, \nu) $ as $ \nu \to \infty $, equal to $ B_0 $.

Moreover, for every fixed $B_0$,  $|B_0|\le \bar B_0=\frac{\sqrt{2}}{4}$ there are two threshold values of the intensity factor of electron-ion collisions $ \nu_1 (B_0) $ and $ \nu_2 (B_0) $, $ \, 0 <\nu_1 (B_0) <\nu_2 (B_0) <\infty $ such that for $ \nu \in [\nu_1, \nu_2] $ the oscillations of the medium are completely suppressed. For other relationships between the parameters $ B_0 $ and $ \nu $, including $ |B_0| \ge \bar B_0 $, stabilization is accompanied by oscillations.

As for the criterion for the formation of singularities, similar to the previous cases, it can also be obtained. However, it is technically much more complex due to the fact that there are five independent parameters. The analytical formulas obtained here are very cumbersome, which hampers the analysis of the solution. However, the use of numerical methods allows one to study the domain of smoothness of the solution when some of the parameters are fixed, and using the methods of computer algebra, one can study the limiting behavior of the oscillation frequency.

The paper is organized as follows. In Sec.\ref{S2} we obtain a matrix Riccati equation to describe the behavior of the space derivatives of the solution. Then we use the Radon lemma to linearize it and obtain an implicit criterion of the singularity formation. In Sec.\ref{S3}
we propose an algorithm to find the domain of smoothness of the solution numerically and test it for the initial data corresponding to a standard laser pulse, traditionally used in numerical computations. In Sec.\ref{S4} we study the character of stabilization of a smooth solution, in particular, oscillations, induced by a large intensity of collisions. In Sec.\ref{S5} we discuss the results and their applicability in physics.

\section{A criterion of the singularity formation}	\label{S2}

System \eqref{3gl5non} along characteristics can be written as follows:
\begin{eqnarray}\label{charV}
\frac{dV_1}{d \theta}=-  E- B_0V_2-\nu V_1,\quad
\frac{dV_2}{d \theta}= B_0V_1-\nu V_2,\quad%\\\nonumber
\frac{d E}{d \theta}= V_1,\quad
\frac{d\rho}{d \theta}= V_1.
\end{eqnarray}
It immediately implies that along every characteristic $V_1^2(t)+V_2^2(t)+E^2(t)\le V_1^2(0)+V_2^2(0)+E^2(0)$, therefore components of the solution remain bounded till the moment of a possible singularity formation.

Let us denote $q_1= \frac{\partial V_1 }{\partial \rho}, q_2= \frac{\partial V_2 }{\partial \rho}, s= \frac{\partial  E}{\partial \rho}$ and \eqref{3gl5non} with respect to $\rho$ and obtain along characteristics $\frac{d\rho}{d \theta}= V_1$
the following Cauchy problem:
\begin{equation}
\label{ch}
 \frac{d q_1}{d \theta}=-q_1^2-s-B_0q_2-\nu q_1,
   \quad
 \frac{d q_2}{\partial \theta}=-q_1q_2+B_0q_1-\nu q_2,
   \quad
   \frac{d s}{d \theta}=q_1(1-s),
\end{equation}

\begin{equation}\label{cd222}
     q_1(\rho,0) = q_1^0(\rho), \quad q_2(\rho,0) = q_2^0(\rho), \quad
     s(\rho,0) = s^0(\rho).
\end{equation}
Note that \eqref {ch} is split off from the \eqref {charV} system and can be written as the matrix Riccati equation.

In what follows we need the following version of the Radon lemma (1927) \cite{Riccati}, Theorem 3.1, see also \cite{Radon}.

\begin{theorem}[The Radon lemma]
\label{T2}
A matrix Riccati equation
\begin{equation}
\label{Ric}
 \dot W =M_{21}(t) +M_{22}(t)  W - W M_{11}(t) - W M_{12}(t) W,
\end{equation}
 {\rm (}$W=W(t)$ is a matrix $(n\times m)$, $M_{21}$ is a matrix $(n\times m)$, $M_{22}$ is a matrix  $(m\times m)$, $M_{11}$ is a matrix  $(n\times n)$, $M_{12} $ is a matrix $(m\times n)${\rm )} is equivalent to the homogeneous linear matrix equation
\begin{equation}
\label{Lin}
 \dot Y =M(t) Y, \quad M=\left(\begin{array}{cc}M_{11}
 & M_{12}\\ M_{21}
 & M_{22}
  \end{array}\right),
\end{equation}
 {\rm (}$Y=Y(t)$  is a matrix $(n\times (n+m))$, $M$ is a matrix $((n+m)\times (n+m))$ {\rm )} in the following sense.

Let on some interval ${\mathcal J} \in \mathbb R$ the matrix-function $\,Y(t)=\left(\begin{array}{c}Q(t)\\ P(t)
  \end{array}\right)$ {\rm (}$Q$  is a matrix $(n\times n)$, $P$  is a matrix $(n\times m)${\rm ) } be a solution of \eqref{Lin}
  with the initial data
  \begin{equation}\label{LinID}
  Y(0)=\left(\begin{array}{c}I\\ W_0
  \end{array}\right)
  \end{equation}
   {\rm (}$ I $ is the identity matrix $(n\times n)$, $W_0$ is a constant matrix $(n\times m)${\rm ) } and  $\det Q\ne 0$ on ${\mathcal J}$.
  Then
{\bf $ W(t)=P(t) Q^{-1}(t)$}
is the solution of \eqref{Ric} with $W(0)=W_0$ on ${\mathcal J}$.
\end{theorem}

In our case \eqref{ch}
%$$
\begin{eqnarray}\label{M}
W=\begin{pmatrix}
  q_1\\
  q_2 \\
  s
\end{pmatrix}, M_{21}=\begin{pmatrix}
  0\\
  0 \\
  0
\end{pmatrix}, M_{22}=\begin{pmatrix}
  - \nu & -B_0 & -1\\
  B_0 & - \nu & 0\\
  1 & 0 & 0\\
\end{pmatrix},\\\nonumber
%$$ $$
M_{11}=\begin{pmatrix}
  0\\
\end{pmatrix}, M_{21}=\begin{pmatrix}
 1 & 0 & 0\\
\end{pmatrix}.
\end{eqnarray}
%$$

\begin{lem}\label{L1} The eigenvalues of matrix $M$, constructed by the rules of Theorem \ref{T2} for the blocks \eqref{M}
are the following:  $\lambda_1=0$,  $\lambda_i$, $i=2,3,4$ are roots of a 3rd order algebraic equation
\begin{equation}\label{eqlam}
\lambda^3+2\nu \lambda^2+(1+B_0^2+\nu^2)\lambda+\nu=0.
\end{equation}
The real parts of $\lambda_k$, $k=2,3,4$  are negative for $\nu>0$.
\end{lem}
\proof
The method for obtaining \eqref{eqlam} is standard. We use the Routh table for polynomials to show that all roots of \eqref{eqlam} are in the left half-plane. $\Box$

\begin{lem}\label{L2}
The solution to the Cauchy problem
\eqref{Lin}, \eqref{LinID} for the matrix $M$, constructed by the rules of Theorem \ref{T2} for the blocks \eqref{M}, is
\begin{equation*}%\label{eqlam}
W(\theta)=\sum\limits_{k=1}^4 C_k \bv_k e^{\lambda_k \theta},
\end{equation*}
where $\bv_k=v_{kj}$, $k,j=1,\dots, 4$ are the eigenvectors of $M$ and $C_k$, $k=1,\dots, 4$ are solutions to the algebraic linear system
\begin{equation*}%\label{eqlam}
W(0)=\sum\limits_{k=1}^4 C_k \bv_k.
\end{equation*}
The function $Q(\theta)$, a part of the solution, has the form
\begin{equation}\label{Q}
Q(\theta)=\sum\limits_{k=1}^4 C_k v_{k1} e^{\lambda_k \theta}.
\end{equation}
\end{lem}
The {\it proof} is standard. The function $Q(\theta)$ depends on the initial data $   q_1^0,q_2^0,s^0$ and parameters $B_0, \nu$.

\begin{theorem}\label{T3}
For the existence of a $ C^1 $ -- smooth solution  $ (V_1 (\theta, \rho), \, V_2 (\theta, \rho), \, E (\theta, \rho)) $ of the problem  (\ref {3gl5non}), (\ref {cd2}), where $(V_1^0, V_2^0, E^0)\in C^2({\mathbb R}) $,
it is necessary and sufficient that at any point $ \rho \in \mathbb R $ the function $Q(\theta)$, given as \eqref{Q}, does not have roots
at the half-axis $\theta>0$.
Otherwise, the solution blows up at the time
\begin{equation}\label{Tc}
  T_c=\inf\limits_{\theta_*>0} \,\{\rho_0 \in  {\mathbb R} \, | \,  Q(\theta_*, \rho_0)=0\}.
 \end{equation}
\end{theorem}

\proof
If we solve the respective system \eqref{Lin}, then we find the solution to
 \eqref{ch} according to Theorem \ref{T2}. The relation
 $ W(\theta)={P(\theta)}Q^{-1}(\theta)$
implies that the derivatives of the solution to the Cauchy problem \eqref{ch}, \eqref{cd222} go to infinity in a finite time along the characteristic starting from the point $ \rho_0 \in \mathbb R $ if and only if
  there is $ \theta_*> 0 $ such that $ Q (\theta_*, \rho_0) = 0 $. The smallest root $ \theta_*> 0 $ over all  $ \rho_0 \in \mathbb R $ corresponds to the blow up time \eqref{Tc}. $\Box$

In what follows we need  explicit expressions for $\lambda_k$ and $\bv_k$. To write them
we introduce the following constants:
\begin{eqnarray}
K&=&B_0^6+ (2\nu^2+3)B_0^4+(\nu^4-5\nu^2+3)B_0^2+\left(1-\frac34 \nu^2\right),\label{K}\\
K_1& =& \left(8\nu^3+36(2B_0^2-1)\nu+24 \sqrt{3 K}\right)^{\frac13},\quad  K_2 = \frac19(3+3 B_0^2-\nu^2).\label{K1K2}
\end{eqnarray}
Then the eigenvalues can be written as
\begin{eqnarray*}
\lambda_2 = \frac16 K_1-6\frac{K_2}{K_1}-\frac23\nu, \quad
\lambda_{3,4}=-\frac{1}{12} K_1+3\frac{K_2}{K_1}-\frac23\nu\pm {\,\rm i \,}\frac{\sqrt{3}}{2}
\left(\frac16 K_1+6\frac{K_2}{K_1}\right).
\end{eqnarray*}
The respective eigenvectors are
\begin{eqnarray*}
\bv_1=\begin{pmatrix}
  1\\
  0 \\
  0 \\
  0
\end{pmatrix},\quad
\bv_k=\begin{pmatrix}
   -\frac{\nu}{((\lambda_k+\nu)^2+1+B_0^2)\lambda_k}\\
   -\frac{\nu}{(\lambda_k+\nu)^2+1+B_0^2} \\
  -\frac{\lambda_k^2+\lambda_k\nu+1+B_0^2}{B_0((\lambda_k+\nu)^2+1+B_0^2)}\\
  1
\end{pmatrix},\quad k=2,3,4.
\end{eqnarray*}

\section{Global smoothness induced by  the regularizing factors for a standard pulse}	\label{S3}

We call the "standard pulse" a set of initial data that looks like:
%We call a "standard impulse" a dataset  this:
\begin{equation}
V_1^0(\rho) = V_2^0(\rho)= 0,\quad E_0(\rho) = k\, \rho \,\exp\left\{-
\dfrac{\rho^2}{\sigma}\right\}, \quad k = \left(\dfrac{a_*}{\rho_*}\right)^2, \; \sigma=\frac{\rho_*^2}{2}.
\label{ddd}
\end{equation}
The data ~(\ref{ddd}) was chosen in the assumption that the oscillations are excited  by a laser pulse with a frequency of $ \zo_l $ ($ \zo_l \gg \zo_p $) when it is focused in a line, which can be achieved by using a cylindrical lens~\cite{Shep13}.
The parameters $ \rho_* $ and $ a_* $ characterize the scale of the localization region and the maximum value
$ E_{\max} = a_*^2 / (\rho_* 2 \sqrt {{\rm e}}) \approx 0.3 a_*^ 2 / \rho_* $ of the electric field, respectively. The specific values of $ \rho_* $ and $ a_* $ are not important here. This kind of data are traditionally used for numerical simulation of different phenomena in  a rarefied plasma, see e.g. \cite{CH18}.

Although the way of applying Theorem \ref {T3} is formally clear, the technical side is fraught with great difficulties.
Indeed, we need to solve the transcendental equation $ Q (\theta) = 0 $ with very cumbersome coefficients, which depends on the eigenvectors of $M$.
%We do not write the explicit form of $Q(t)$ here.
Nevertheless, for any fixed set of parameters $   q_1^0, q_2^0, s^0, B_0, \nu$ we use a numerical procedure to study the effect of each particular parameter.

To study the blow-up for the data \eqref{ddd} we choose the most "dangerous" point $\rho_0=0$, where the derivative $(E_0)'$ has a maximum. There $   q_1^0=q_2^0=0, \, s^0=k$, this reduces the number of free parameters to three and simplifies the analysis.

First, we picture on the plane $(\nu, B_0)$ the set of parameters  that leads to a finite time blow-up for a fixed $s^0=k$.
If $k<\frac12$, then this set is empty. This follows, for example, from Theorem \ref{T1} for the case $\nu=0$ and from the estimate
of the critical $k$ from \cite{RChD20}, $\nu\in [0,2)$, $B_0=0$,
\begin{equation}\label{k_kr}
k_{cr}=\frac{1}{1+\exp(-\frac{\nu \pi}{\sqrt{4-\nu^2}})}.
\end{equation}
We note also that the nonnegativity of density dictates the requirement $k<1$ (see \eqref{3gl4den}). Thus, we choose for the numerical illustrations the interval $k=s^0\in (\frac12,1)$. To estimate the sign of  minimum of $Q(\theta)$ for $\theta>0$  we used
Newton's method built into the Optimization package of MAPLE.
Fig.1,  presents the "blow-up" domains for $k=0.6$, $k=0.7$ and $k=0.8$ and shows that this domain expands with increasing $ k $.
The limits of $ \nu $ as $ B_0 \to 0 $ on the boundary of the domain correspond to \eqref{k_kr}.
\begin{center}
\begin{figure}[htb]
%\hspace{-1.5cm}
%\begin{minipage}{0.4\columnwidth}
\centerline{
\includegraphics[scale=0.4]{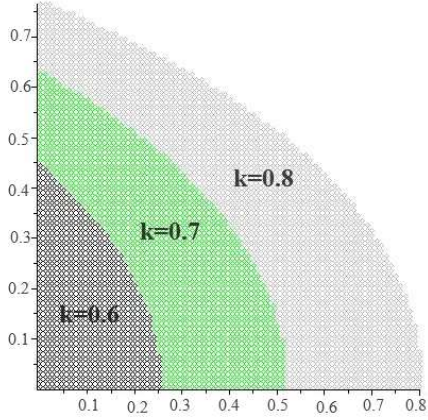}}
%\vspace{-0.5 cm}
%\end{minipage}
%\hspace{1cm}
%\begin{minipage}{0.4\columnwidth}
%\centerline{
%\includegraphics[scale=0.3]{Pic2.1.ps}}
%\vspace{-0.5 cm}
%\end{minipage}
\caption{ The "blow up" domains on the plane $  \nu $  (horizontal) and $ B_0$ (vertical) for different values of $k$, the step is 0.01 for both variables. The picture is symmetrical about the axis $B_0=0$.}\label{Pic1}
\end{figure}
\end{center}
The next numerical experiment that we can perform is to fix $\nu$ and $B_0$ and to find a domain $\mathcal D$, consisting of points $q_1^0, q_2^0, s^0$, corresponding to a globally in time smooth solution \eqref{3gl5non}, \eqref{ddd}, provided at every point $\rho\in \mathbb R$ the derivatives of the initial data fall in this domain. In other words, we try to picture an analog of the plain domain found in \cite{RChD20} for a fixed $\nu$. However, the considered domain $ \mathcal D $ is three-dimensional, and the result of calculations is difficult to present in a convenient form. Therefore one can draw a section of this domain, for example, with $q_2^0=\rm const$. We do not present  the result of this computation here, just note that the "blow up" domain, the complement of the domain $\mathcal D$ to the entire space, expands with  increasing  $q_2^0$.

\section{Character of stabilization}	\label{S4}

In this section, we establish certain properties of smooth solutions.

\begin{theorem}\label{T4}
Let the initial data \eqref{cd2} are such that the solution of the Cauchy problem \eqref{3gl5non}, \eqref{cd2}, %$\nu>0$
is globally $C^1$ -- smooth.

1. For any fixed $B_0$ there  exists  $\lim\limits_{\nu\to\infty}\Im\, \lambda_3 (B_0, \nu) =B_0 $;

2. The motion is oscillatory if and only if $K>0$, given as \eqref{K}, is positive. If, in addition, $\nu>0$, then the solution decays
 to the trivial steady state. If $K\le 0$, then the solution  decays to the trivial steady state monotonically starting from a sufficiently large $\theta$.

\end{theorem}
%for any fixed $B_0$ there exists a limit of the frequency $\Omega (B_0, \nu)$ of oscillations as $\nu\to \infty$, equal to $B_0$.
\proof
1. The first properties follows from the expansion $$\lambda_3 (B_0, \nu)=-\nu+{\,\rm i \,}B_0\,  +O\left(\frac{1}{\nu}\right), \quad\nu\to\infty.$$
It implies that the oscillations cannot be suppressed by a large frequency of collisions.

2.
If the function $Q(\theta)$ does not have roots for $\theta>0$ for all $\rho_0\in \mathbb R$, then taking into account Lemma \ref{L2} we conclude that the solution to \eqref{3gl5non}, \eqref{cd2}
stabilizes to  to the trivial state provided $\nu>0$ since $e^{\lambda_k\theta}\to 0$, $\theta \to \infty$, $k=2,3,4$.

The stabilization is monotonic starting from a certain moment of time if and only if $\Im \lambda_k=0$ for all $k=2,3,4.$

If $K$ is nonnegative, then  $\Im\, \lambda_2=0$, and  $\Im\, \lambda_{3,4}\ne 0$ provided $K_1^2+36 K_2\ne 0$.
 It can be readily checked that $K_1^2+36 K_2= 0$ implies $K=0$.
 We take into account that $K_1$ and $K_2$, given as \eqref{K1K2}, both vanish at the hyperbola  $3B_0^2-\nu^2+9=0$, so the ratio $\frac{K_2}{K_1}$ is finite.

If $K< 0$, then it suffices to prove that $\Im\, \lambda_3=0$. Since $|\Im\, \lambda_3|=|\Im\, \lambda_4|$,  Lemma \ref{L1} implies that
$\lambda_k$, $k=2, 3, 4$, are real and negative.
First of all we notice that if $K<0$, then necessarily $K_1=a+{\,\rm i\,}b\, $, $a, b\in \mathbb R$, $ab\ne 0$. Then it is easy to compute that
\begin{eqnarray}\label{Il3}
\Im\, \lambda_3=\frac{\sqrt{3}a-b}{12(a^2+b^2)} (36 K_2+a^2+b^2).
\end{eqnarray}
Since $K_1= (A\pm{\,\rm i\,}B)^\frac13,$ $A=
8\nu^3+36(2B_0^2-1)\nu,$ \,$B=24 \sqrt{-3K}$, we can find $a$ and $b$,  then substitute the result in \eqref{Il3} and check directly that
$\Im\, \lambda_3=0$. This is a rather cumbersome computation, however, it can be done using computer algebra.

Note that since we are extracting the 3rd order roots of complex numbers, we must
account for the corresponding branch of this multivalued function.
$\Box$

\begin{corollary} Let the solution of \eqref{3gl5non}, \eqref{cd2} is globally smooth in time.
For every fixed $B_0$,  $|B_0|\le \bar B_0=\frac{\sqrt{2}}{4}$ there are two threshold values of the factor of the intensity of collisions, $ \nu_1 (B_0) $ and $ \nu_2 (B_0) $, $ \, \frac34 \sqrt{6} <\nu_1 (B_0) <\nu_2 (B_0) <+\infty $ such that for $ \nu \in [\nu_1, \nu_2] $ the oscillations of the medium are completely suppressed. For other relations between the parameters $ |B_0| $ and $ \nu>0 $, including $B_0\ge \bar B_0$, the stabilization is accompanied by oscillations.
\end{corollary}

For the {\it proof }  we have to study the curve $K(B_0,\nu)=0$ on the plane $(B_0, \nu)$.
 We resolve $K=0$ with respect to $\nu$ and get two branches $\nu=\nu_\pm(B_0)$ with the following properties:
\begin{eqnarray*}
\nu_\pm=\frac14 \frac{\sqrt{2+40 B_0^2-16 B_0^4\pm 2\sqrt{1-24 B_0^2+192 B_0^4-512 B_0^6}}}{B0}.
\end{eqnarray*}
 Thus,
 \begin{eqnarray*}
 \quad B_0\in \left[-\bar B_0, \bar B_0\right],\quad
 \bar B_0=\frac{\sqrt{2}}{4}\approx 0.353\dots,\quad
 \nu_\pm \left(\bar B_0\right)=\frac34 \sqrt{6}\approx  1.837\dots,
 \end{eqnarray*}
 the limit values are
  \begin{eqnarray*}
\lim\limits_{B_0\to 0} \nu_-=2,\quad \lim\limits_{B_0\to 0} \nu_+=+\infty.
\end{eqnarray*}
The points $K(B_0,\nu)=0$ for $B_0\ge 0$ are presented in Fig.2,
%\ref{Pic3},
 the picture is symmetric with respect to $B_0=0$.
Thus, if we fix any $B_0\in \left[-\bar B_0, \bar B_0\right]$ and begin to increase $\nu$, we start in the domain $K>0$, then at $\nu_1=\nu_-(B_0)$ we fall in the domain $K\le 0$ and then leave this domain at $\nu_2=\nu_+(B_0)$. This observation proves the corollary.
We note that the results is in compliance with the case $B_0=0$, studied in \cite{RChD20}, where it was shown that the monotonic decay takes place for $\nu\ge 2$.
$\Box$
\begin{center}
\begin{figure}[htb]
\hspace{-1.5cm}
%\begin{minipage}{0.4\columnwidth}
%\centerline{
\includegraphics[scale=0.5]{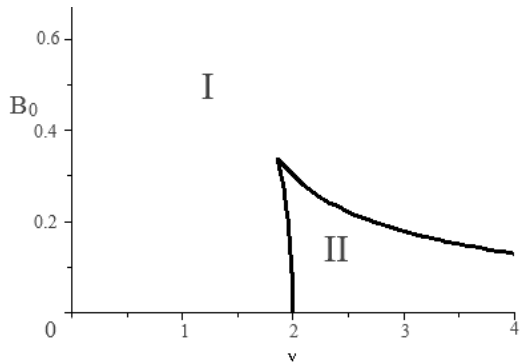}
%\vspace{-0.5 cm}
%\end{minipage}
\caption{Values $ B_0, \, \nu $ corresponding to oscillatory regime (I) and monotonic regime (II)  for $B_0\ge 0$; the picture is symmetrical about the axis $B_0=0$. The "blow up" domain for the standard pulse (Sec.\ref{S3}, Fig.1) is located near the origin of coordinates and belongs to the oscillatory regime.  }\label{Pic3}
\end{figure}
\end{center}

\begin{remark}
The following expansions also make it possible to trace the effect of low collision intensity on the character of oscillations:
 \begin{eqnarray*}
 \lambda_{2} (B_0, \nu)&=& - \frac{1}{B_0^2+1} \nu+O\left(\nu^2\right), \quad \nu\to 0,\\
 \lambda_{3,4} (B_0, \nu)&=&\pm {\,\rm i\,}\sqrt{B_0^2+1}- \frac12 \frac{2B_0^2+1}{B_0^2+1} \nu +O\left(\nu^2\right).
 \end{eqnarray*}
We see that in the first approximation, the low intensity of collisions does not change the frequency of the oscillations, but causes their exponential decay. Generally speaking, the rate of this damping does not decrease at large values of $|B_0|$.
 \end{remark}
\begin{remark}
It should be noticed that the oscillations induced by friction are known in mechanical systems (see, e.g. \cite{Bigoni}).
\end{remark}

\section{Discussion}	\label{S5}

We performed a mathematical analysis of equations of upper hybrid oscillations \eqref{3gl5non} for all allowed values of the constant parameters $\nu$ and $B_0$. Nevertheless, it should be noted that the most interesting phenomena, such as the persistence  of oscillations at large values of $ \nu $, can hardly be observed experimentally, since the physical values of $\nu$ are of the order  $ 10^{-2} $ (see  \cite{RChD20} for details).  For small values of $ \nu $, the effect of the external magnetic field is very close to the case $ \nu = 0 $ considered in \cite {bib:03}: basically, increasing the magnetic field prevents blow-up (except for specially adapted initial data).
It should also be noted that the most powerful regularizing factor is the dependence of the intensity of electron-ion collisions on the density. An analysis of this situation in the absence of a magnetic field was carried out in \cite{R2022}. It is natural to expect that, in the presence of a magnetic field, the effect of suppressing the formation of singularities for all smooth initial data will be retained.

\section*{Acknowledgment}
Supported by the Ministry of Education and Science of the Russian Federation as part of the program of the Moscow Center for Fundamental and Applied Mathematics under the agreement 075-15-2019-1621.

The authors are grateful to E.V.Chizhonkov for drawing attention to the problem, stimulating discussions and constant interest.

\end{document}